\documentclass[final,3p,times,twocolumn]{elsarticle}

\usepackage{graphics}
\usepackage{amssymb}
\usepackage{amssymb}
\usepackage[latin1]{inputenc}
\usepackage{amsmath}
\usepackage{amsfonts}
\usepackage{bm}
\usepackage{graphicx}
\usepackage{epsfig}
\usepackage{subfigure}
\usepackage{setspace}%

\journal{Journal of Luminescence}

\begin{document}

\begin{frontmatter}

\title{Efficient optical pumping of Zeeman spin levels in Nd$^{3+}:$YVO$_{4}$}

\author{Mikael Afzelius \corref{cor1}}
\ead{mikael.afzelius@unige.ch}
\cortext[cor1]{Corresponding author}
\author{Matthias U. Staudt\corref{cor2}}
\author{Hugues de Riedmatten}
\author{Nicolas Gisin}
\address{Group of Applied Physics, University of Geneva, CH-1211 Geneva 4, Switzerland}
\author{Olivier Guillot-Noël}
\author{Philippe Goldner}
\author{Robert Marino}
\author{Pierre Porcher}
\address{Ecole Nationale Supérieure de Chimie de Paris (ENSCP), Laboratoire de Chimie
de la Matière Condensée de Paris, CNRS-UMR 7574, ENSCP, 11 rue Pierre et Marie
Curie 75231 Paris Cedex 05, France}
\author{Enrico Cavalli}
\address{Dipartimento di Chimica Generale ed Inorganica, Chimica Analitica e Chimica,
Fisica, Università di Parma, Viale G. P. Usberti 17/a, 43100 Parma, Italy.}
\author{Marco Bettinelli}
\address{Dipartimento Scientifico e Tecnologico, Univ. Verona Strada Le Grazie 15,
37134 Verona, Italy}

\begin{abstract}
We demonstrate that Zeeman ground-state spin levels in Nd$^{3+}$:YVO$_{4}$ provides the possibility to create an efficient $\Lambda$-system for optical pumping experiments. The branching ratio $R$ in the $\Lambda$-system is measured experimentally via absorption spectroscopy and is compared to a theoretical model. We show that $R$ can be tuned by changing the orientation of the magnetic field. These results are applied to optical pumping experiments, where significant improvement is obtained compared to previous experiments in this system. The tunability of the branching ratio in combination with its good coherence properties and the high oscillator strength makes Nd$^{3+}:$YVO$_{4}$ an interesting candidate for various quantum information protocols.
\end{abstract}


\end{frontmatter}

%% \linenumbers

%% main text
\section{Introduction}

Rare-earth (RE) ion doped solids have recently gained interest in the context of quantum information processing applications, particularly as storage devices for light at the single photon level \cite{Hammerer2008}. These quantum memories \cite{Julsgaard2004,Chaneliere2005,Eisaman2005,Choi2008,deRiedmatten2008} have a role in future quantum networks \cite{Kimble2008}, particularly in quantum repeaters \cite{Briegel1998,Duan2001} that can provide a scalable solution for long-distance quantum communication. In that context RE ion doped solids are interesting due to their long coherence times at cryogenic temperatures \cite{Macfarlane2002}, which could provide coherent storage of single photons on long time scales.

In quantum memory schemes based on electromagnetically induced transparency
(EIT) \cite{Fleischhauer2000,Lukin2003}, controlled reversible inhomogenenous broadening (CRIB) \cite{Moiseev2001,Kraus2006,hetet2008} or atomic frequency combs (AFC) \cite{afzelius08}, the system is initialized by preparing the atoms in a well-defined ground state (state initialization). This is often realized by
the usage of a $\Lambda$-system, which consists of two ground state levels
(typically two spin levels) that are optically connected to one excited state.
The state initialization (or spin polarization) is done by optically pumping
atoms to the excited state via one leg of the $\Lambda$-system and then
waiting for the atoms to de-excite to the ground state via the other leg of
the $\Lambda$-system. The capability to perform good spin polarization, with close to 100\% of the spins being transferred to one of the spin states, is crucial to all quantum memory schemes. The efficiency of the spin polarization depends on several parameters; (1) on the ground state and excited state lifetimes and (2) on the branching ratio $R$, which describes the probability of an excited ion to change spin level while decaying. Good branching ratios and long-lived ground states can be found in non-Kramers ions as praseodymium \cite{Holliday1993,Equall1995}, europium \cite{Babbitt1989,Konz2003}, and thulium \cite{Ohlsson2003,gui05,louchet08}. It has been considerably more difficult to obtain similar results in Kramers ions such as erbium \cite{Hastings2008a,lauritzen08} and neodymium \cite{Hastings2008}, due to short-lived ground states and poor branching ratios. Yet, these Kramers ions are still interesting as candidates for quantum memories due to their long optical coherence times \cite{Sun2002} and practical wavelengths where diode lasers can be used.\\
\indent Here we demonstrate that Zeeman ground-state spin levels in Nd$^{3+}:$%
YVO$_{4}$ offers a magnetically tunable branching ratio $R$, resulting in an efficient $\Lambda$-system. The branching ratio is found experimentally by measuring the inhomogeneous absorption profile in a moderate magnetic field of 0.3 T, where the relative probabilities of the Zeeman transitions can be directly measured from the
inhomogeneously broadened absorption spectrum. The branching ratio is measured
for two magnetic field orientations and the results are compared to calculations based on crystal field theory and an effective spin $\tilde{S}=1/2$ model. Using the found branching value we experimentally demonstrate
efficient spin initialization ($97.5\%$ spin polarization) by optical pumping
in the $\Lambda$-system. These results are interesting in view of the good optical coherence times found in Nd$^{3+}:$%
YVO$_{4}$ \cite{Hastings2008,Sun2002} and the recent demonstration of light storage at the single photon level in this system \cite{deRiedmatten2008}.

\section{Theory}
\label{sec_theory}
\begin{figure}[ptb]
\begin{center}
\includegraphics[width=.55\textwidth]{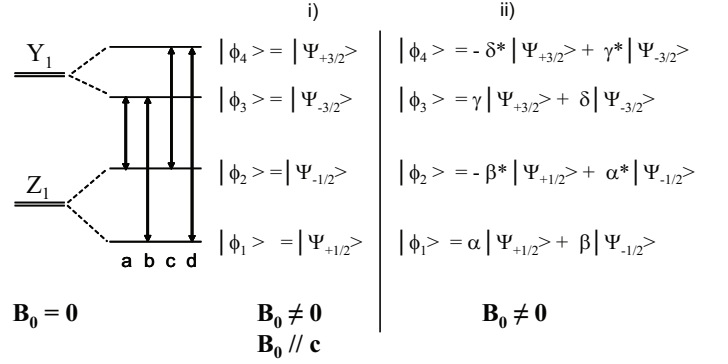}
\caption{Investigated transitions, levels and wavefunctions of
  Nd$^{3+}$ for  a magnetic field $\mathbf{B}_0$ along the c-axis
  (i) or at an arbitrary angle $\theta$ (ii).}
\label{Fig1}
\end{center}
\end{figure}

We here investigate the $^{4}$I$_{9/2}\rightarrow$ $^{4}$F$_{3/2}$ transition
at 879 nm of Nd$^{3+}$ ions doped into a YVO$_{4}$ crystal. This crystal is
uniaxial, and Nd$^{3+}$ ions substitute for Y$^{3+}$ ions in sites of D$_{2d}$
point symmetry. In a crystal field of this symmetry, the ground state $^{4}%
$I$_{9/2}$ splits into five Kramers doublets (Z$_{1}$ to Z$_{5}$) and the
$^{4}$F$_{3/2}$ excited state splits into two Kramers doublets (Y$_{1}$ and
Y$_{2}$). The transition we consider here is between Z$_{1}$ and Y$_{1}$, see
Fig. \ref{Fig1}. Under an applied magnetic field the ground and excited state
doublets split into two spin sub-levels via a first-order Zeeman interaction.
This gives rise to a four-level system with two ground levels and two excited
levels, and the associated four possible transitions hereafter denoted a,b,c and d (see Fig. \ref{Fig1}).

The calculation of the branching ratios between the different
transitions is based on the electronic wavefunctions of Nd$^{3+}$ ions
in YVO$_{4}$ obtained from the diagonalization of the free-ion and
crystal-field Hamiltonians, using the parameters obtained by Guillot-Noël et al. in
Ref. \cite{ogn98}. This calculation shows that for the
$^{4}$I$_{9/2}$ (Z$_{1}$) and for the $^{4}$F$_{3/2}$ (Y$_{1}$)
Kramers' doublets, the electronic wavefunctions are of the form
(neglecting term
and $J-$mixing):
\begin{eqnarray}
  \label{eq:1}
\lambda_1\left\vert ^{4}I_{9/2}\text{,
  }M=\pm\frac{1}{2}%
\right\rangle +\lambda_2\left\vert ^{4}I_{9/2}\text{, }M=\mp\frac{7}%
  {2}\right\rangle
  \\ \nonumber
  +\lambda_3\left\vert ^{4}I_{9/2}\text{,
  }M=\pm\frac{9}%
  {2}\right\rangle
\end{eqnarray}
  and
  \begin{eqnarray}
    \label{eq:2}
 \left\vert ^{4}F_{3/2}\text{, }M=\pm\frac{3}%
  {2}\right\rangle
  \end{eqnarray}
 where $M$ is the projection of the total $J$
moment and the $\lambda_i$ are complex coefficients. In a
D$_{2d}$ site symmetry, for a $^{2S+1}L_{J}$ multiplet, the
wavefunctions are a sum of states $ \left|^{2S+1}L_{J},\mu \pm
  4m\right>$ where $\mu$ is the crystal field quantum
number \cite{hellwege49} and $m$ an integer such that $-J\leq \mu
\pm4m\leq J$.  For the $^{4}$I$_{9/2}$ (Z$_{1}$), we have $\mu=\pm1/2$
and for the $^{4}$F$_{3/2}$ (Y$_{1}$), $\mu=\pm 3/2$ as can be seen
from Eqs. \ref{eq:1} and \ref{eq:2}.
Without an external magnetic field, the states characterized by $+\mu$ and
$-\mu$ are degenerate because of time reversal symmetry (Kramers'
theorem).

  Under an external magnetic field $\mathbf{B}_{0}$, the
electronic Zeeman interaction Hamiltonian $H_{EZ}$ has to be added to
the free ion and crystal field Hamiltonians, where $H_{EZ}$ is expressed
as \cite{Wybourne65}%
\begin{equation}
H_{EZ}=\beta\mathbf{B}_{0}\mathbf{\cdot}\sum_{i=1}^{n}\left(  \mathbf{l}%
_{i}+g_{s}\mathbf{l}_{i}\right)  =\beta\mathbf{B}_{0}\mathbf{\cdot}\left(
\mathbf{L}+g_{s}\mathbf{S}\right).  \label{EZ}%
\end{equation}
Here $\beta$ is the electronic Bohr magneton, $\mathbf{l}_{i}$ and
$\mathbf{s}_{i}$ are the individual orbital and spin momentum of the electrons
and $\mathbf{L=}\sum\limits_{_{i=1}}^{n}\mathbf{l}_{i}$ et $\mathbf{S=}%
\sum\limits_{_{i=1}}^{n}\mathbf{s}_{i}$. $g_{s}$ is the gyromagnetic ratio of
the electron spin which is equal to 2.0023. Under an external magnetic field,
each Kramers' doublet split into two electronic singlets.

From the crystal field calculations, when $\mathbf{B}_{0}$ is parallel
to the c-axis, the wavefunctions of the splitted levels are given by
$\left\vert \Psi_{+\frac{1}{2}}\right\rangle $ and $\left\vert
  \Psi_{-\frac{1}{2}%
  }\right\rangle $ for the lowest and the highest level of the
$^{4}$I$_{9/2}$ (Z$_{1}$) doublet and by $\left\vert
  \Psi_{-\frac{3}{2}}\right\rangle $ and $\left\vert
  \Psi_{+\frac{3}{2}}\right\rangle $ for the lowest and the highest
level of the $^{4}$F$_{3/2}$ (Y$_{1}$) doublet (Fig. \ref{Fig1}.i). A
general orientation of the external magnetic field $\mathbf{B}_{0}$,
mixes the $\left\vert \Psi_{\mu }\right\rangle $ and $\left\vert
  \Psi_{-\mu}\right\rangle $ states  and we have
for the $^{4}$I$_{9/2}$ (Z$_{1}$) doublet two splitted levels
$\left\vert \phi_{1}\right\rangle $ (low energy) and $\left\vert
  \phi_{2}\right\rangle $ (high energy) given by (Fig.  \ref{Fig1}.ii):%
\begin{align}
\label{eq:9}
\left\vert \phi_{1}\right\rangle  &  =\alpha\left\vert \Psi_{+\frac{1}{2}%
}\right\rangle +\beta\left\vert \Psi_{-\frac{1}{2}}\right\rangle \\
\left\vert \phi_{2}\right\rangle  &  =-\beta^{\ast}\left\vert \Psi_{+\frac{1}{2}%
}\right\rangle +\alpha^{\ast}\left\vert \Psi_{-\frac{1}{2}}\right\rangle \nonumber
\end{align}
and for the $^{4}$F$_{3/2}$ (Y$_{1}$) doublet, two splitted levels $\left\vert
\phi_{3}\right\rangle $ (low energy) and $\left\vert \phi_{4}\right\rangle $ (high
energy) given by:%
\begin{align}
\label{eq:10}
\left\vert \phi_{3}\right\rangle  &  =\gamma\left\vert \Psi_{+\frac{3}{2}%
}\right\rangle +\delta\left\vert \Psi_{-\frac{3}{2}}\right\rangle \\
\left\vert \phi_{4}\right\rangle  &  =-\delta^{\ast}\left\vert \Psi_{+\frac{3}{2}%
}\right\rangle +\gamma^{\ast}\left\vert \Psi_{-\frac{3}{2}}\right\rangle. \nonumber
\end{align}
In the following, two branching ratios will be discussed:
\begin{equation}
  \label{eq:3}
 R_{\parallel}=\frac{\left\vert \left\langle \phi_{2}\right\vert \widehat{P}_{\parallel}\left\vert
\phi_{3}\right\rangle \right\vert ^{2}}{\left\vert \left\langle \phi
_{1}\right\vert \widehat{P}_{\parallel
}\left\vert \phi_{3}\right\rangle \right\vert ^{2}%
}
\end{equation}
and
\begin{equation}
  \label{eq:4}
   R_{\bot}=\frac{\left\vert \left\langle \phi_{2}\right\vert \widehat{P}_{\bot}\left\vert
\phi_{3}\right\rangle \right\vert ^{2}}{\left\vert \left\langle \phi
_{1}\right\vert \widehat{P}_{\bot}\left\vert \phi_{3}\right\rangle \right\vert ^{2}%
}
\end{equation}
where $\widehat{P}_{\parallel}$ is the
transition dipole operator along the c-axis and $\widehat{P}_{\bot}$ is the operator perpendicular
to the c-axis. If we neglect $J$-mixing, the $\Delta J=0,\pm1$
selection rule shows that the $^{4}%
$I$_{9/2}$ (Z$_{1}$)$\rightarrow^{4}$F$_{3/2}$ (Y$_{1}$) transition
has no magnetic dipole component. Thus, $\widehat{P}_{\parallel}$ or
$\widehat{P}_{\bot}$ is an electric dipole operator corresponding to
an electric field polarized parallel ($\mathbf{E}_{\parallel}$) or perpendicular
($\mathbf{E}_{\bot}$) to the c-axis. The weak $J$-mixing appearing in crystal
field calculations could explain some experimental results (see
Section \ref{measure_R}). However, this effect is small and will not be
taken into account in the following.  The selection rules which hold
for electric dipole transitions between the $\left\vert \Psi_{\pm
    \mu}\right\rangle$ states are:
\begin{eqnarray}
  \label{eq:5}
   \left\langle \Psi_{\mu}\right\vert\widehat{P}_{\parallel}\left\vert
\Psi_{\mu'}\right\rangle&=&0\quad \textrm{unless}\quad \mu-\mu'
\quad\textrm{is even}\\
 \left\langle \Psi_{\mu}\right\vert\widehat{P}_{\bot}\left\vert
\Psi_{\mu'}\right\rangle&=&0\quad \textrm{unless}\quad \mu-\mu'
\quad\textrm{is odd}.
\end{eqnarray}
Moreover, the matrix elements for $\widehat{P}_{\parallel}$ or
$\widehat{P}_{\bot}$ are related by:
\begin{align}
  \left\langle \Psi_{\mu}\right\vert \widehat{P}_{\parallel,\bot}\left\vert
    \Psi _{-\mu'}\right\rangle & =-\left\langle \Psi_{-\mu}
  \right\vert \widehat{P}_{\parallel,\bot}\left\vert \Psi_{\mu'}
  \right\rangle\\
  \left\langle \Psi_{\mu}\right\vert \widehat{P}_{\parallel,\bot}
\left\vert \Psi_{\mu'}\right\rangle & =\left\langle \Psi_{-\mu}
\right\vert \widehat{P}_{\parallel,\bot}\left\vert \Psi_{-\mu'}\right\rangle.
\end{align}
These relations can be deduced from the symmetry properties of the
electric dipole operator with
respect to the crystal field and the time reversal
operator \cite{guillot10}.

The branching ratios defined in Eqs. \ref{eq:3} and \ref{eq:4} are
then equal to:%
\begin{equation}
R_{\parallel}=\frac{\left|
      \alpha\gamma-\beta\delta\right|^2}
{\left|\alpha^{\ast}\delta+\beta^{\ast}\gamma \right | ^{2}}\label{branchingratio}%
\end{equation}
and
\begin{equation}
  \label{eq:8}
  R_{\bot}=\frac{\left|
     \beta\gamma+\alpha\delta      \right|^2}
{\left|\alpha^{\ast }\gamma-\beta^{\ast}\delta \right | ^{2}}.
\end{equation}

The $\alpha$, $\beta$, $\gamma$, $\delta$ coefficients are determined
by the orientation of the external magnetic field
$\mathbf{B}_{0}$. They can be calculated directly by diagonalization
of the sum of the free-ion, crystal-field and electronic Zeeman
Hamiltonians or by using an effective
spin-Hamiltonian model. The latter gives more accurate results when
the experimental $g$-factor values are known, as is the case
here. Mehta \textit{et al}. \cite{Mehta00} measured a linear Zeeman effect for
magnetic fields up to 6 T. This implies that at moderate magnetic
field, as the one we use in our experiments (0.3 Tesla), each Kramers'
doublet Z$_{1}$ and Y$_{1}$ can be described by an effective spin
$\widetilde{S}$= 1/2. Quadratic splitting due to the mixing of excited
Kramers' doublets with Z$_{1}$ and Y$_{1}$ are three order of
magnitude smaller than the linear terms. We can thus
perform a first order calculation to determine $\alpha$,
$\beta$, $\gamma$, $\delta$ coefficients by diagonalization of the following
spin-Hamiltonian: \cite{gui05,Mehta00}%
\begin{equation}
H_{Z}=\beta(\mathbf{B}_{0}\cdot g\cdot\mathbf{S})\label{Heffec}%
\end{equation}
where $g$ is the $g$-factor of the doublet under study and
$\mathbf{S}$ the spin 1/2 operator.  Each ground and excited doublet
Z$_{1}$ and Y$_{1}$ is characterized by the principal values $g_\bot$
and $g_{\parallel}$ of the $g$-factor since the latter is axial for a
D$_{2d}$ symmetry. We used the following experimental values
$^{4}$I$_{9/2}($Z$_{1})$:$\left\vert g_{\bot}\right\vert =2.361$,
$\left\vert g_{\parallel}\right\vert =0.915$ \cite{Mehta00};
$^{4}$F$_{3/2}($Y$_{1})$: $\left\vert g_{\bot}\right\vert
=0.28$ \cite{Hastings2008}, $\left\vert g_{\parallel}\right\vert
=1.13$ \cite{Mehta00}. To be consistent with Eqs. \ref{eq:9}
and \ref{eq:10}, $g$-factor values must be negative.  This implies
that the $M_{\widetilde{S}}=-1/2$ spin level is equivalent to the
$\left\vert \Psi_{-\frac{1}{2}%
  }\right\rangle $ state and to the $\left\vert
  \Psi_{+\frac{3}{2}}\right\rangle $ state for the $^{4}$I$_{9/2}$
(Z$_{1}$) and $^{4}$F$_{3/2}$ (Y$_{1}$) doublet, respectively.

\begin{figure}
  \centering
\includegraphics[width=.50\textwidth]{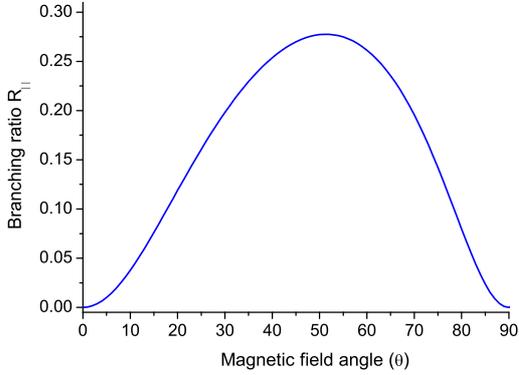}
  \caption{Theoretical $R_\parallel$ branching ratio as a function of the
    angle $\theta$
between the c-axis and the magnetic field. The light is polarized
along the c-axis ($\phi$=0$^{\circ}$).}
\label{fig:Rz_vs_theta}
\end{figure}

In the experiments described below, $\mathbf{B}_0$ lies in the plane
containing the c-axis and the $\mathbf{E}$ electric field. In this
case, all $\alpha$, $\beta$, $\gamma$, $\delta$ coefficients are real,
so that $R_{\parallel}=1/R_\bot$. A plot of $R_\parallel$ as a function of
the magnetic field orientation is presented in
Fig. \ref{fig:Rz_vs_theta}. At $\theta$=0$^{\circ}$ (i.e. $\mathbf{B}_0$ along
the c-axis), Nd$^{3+}$ wavefunctions correspond to pure $ \left|
  \Psi_\mu \right >$ states (Fig. \ref{Fig1}.i). In this case, the
selection rules (Eqs. \ref{eq:5}) show that the a and d
transitions are forbidden for light polarized along the c-axis. Accordingly, the $R_{\parallel}$ ratio vanishes. On the other
hand, when $\mathbf{B}_0$ is directed away from the c-axis,
Nd$^{3+}$ wavefunctions are mixture of $\left| \Psi_\mu \right >$
states and all transitions are allowed (Fig. \ref{Fig1}.ii) until
$\theta=90^\circ$ where the wavefunctions are again pure $\left|
  \Psi_\mu \right >$ states. $R_\parallel$ is predicted to reach its
maximum value, $R_\parallel$=0.278, for $\theta$=51$^{\circ}$ (see Fig. \ref{fig:Rz_vs_theta}). In the experiments presented in the following sections the configuration $\theta$=45$^{\circ}$ was used, where $R_\parallel$=0.270, close to the maximum value.

\section{Experiment}
\label{exp_descr}
The YVO$_{4}$ single crystals doped with Nd$^{3+}$ were grown by spontaneous
nucleation from a Pb$_{2}$V$_{2}$O$_{7}$ flux \cite{Garton1972}. Reagent grade
PbO and V$_{2}$O$_{5}$, Nd$_{2}$O$_{3}$ (99.99 \%) and Y$_{2}$O$_{3}$ (99.99
\%) were used as starting materials in suitable amounts. The doping
concentration was 0.001 \% (Nd/Y nominal molar ratio). The batch was put in a
50 cm$^{3}$ covered platinum crucible and heated to 1300 $^{\circ}$C inside a
horizontal furnace. After a soaking time of 12 h, the temperature was lowered
to 850 $^{\circ}$C at a rate of 3-4 $^{\circ}$C/h, then the crucible was drawn
out from the furnace and quickly inverted to separate the flux from the
crystals grown at the bottom of the crucible. The flux was dissolved by using
hot diluted nitric acid. YVO$_{4}$ crystallizes in the \textit{I41/amd} space
group, with cell parameters a=b=7.118 Å~and c=6.289~Å~and \textit{Z}=4~
\cite{Chakoumakos1994}.

The crystal sample was cooled to a temperature of about 2.8 K in a pulse-tube
cooler. A permanent magnetic field of 310 mT was applied along different
directions. Light at a wavelength of 879 nm excited the $^{4}$I$_{9/2}
\rightarrow$ $^{4}$F$_{3/2}$~transition in the neodymium ions. The light
pulses were created by gating an cw external cavity diode laser (Toptica DL
100) with an acousto-optic modulator (AOM), which was in a double pass
configuration. Pulses of duration 10 ms were created during which the laser
frequency was scanned over about 15 GHz by changing the cavity length using a
piezo. After the AOM the light was split into three beams. A few percent were
coupled into a Fabry-Pérot-cavity for relative frequency calibration during
the scan. Another few percent were used as a reference beam for intensity
normalization, which was detected by a photodiode (Thorlabs PDB150A) before
the crystal. The third beam was focused onto the crystal and was detected in
transmission by a second, identical photodiode. This signal was normalized by
the reference signal in order to account for intensity changes during the
scan. In order to control the polarization of the light incident on the
crystal, a $\lambda/2$-plate was installed directly before the cooler.

\section{Results and Discussion}

\subsection{Measurement of the branching ratio}%
\begin{figure}[ptb]
\includegraphics[width=.50\textwidth]{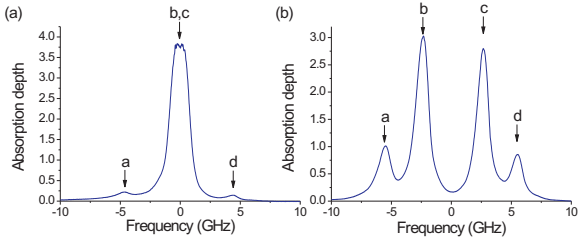}
\caption{Absorption depth as a function of frequency for a magnetic
field (a) parallel $\theta$=0$^{\circ}$ and (b) with an angle of $\theta$=45$^{\circ}$ to the crystal
c-axis. In both measurements the light polarization was chosen to be parallel to the c-axis. The transitions a,b,c,d are marked with the arrows.}%
\label{Fig2}
\end{figure}

\label{measure_R}
The branching ratio $R$ can be determined by measuring the absorption coefficients of the transitions a,b,c and d defined in Fig. \ref{Fig1}. We could directly measure these by applying a relatively strong magnetic field (0.31 T), which splits the Zeeman levels sufficiently so that the different transitions could be resolved, despite the 2 GHz inhomogeneous broadening. We thus measure the transmission spectrum while scanning the laser and calculate the corresponding absorption spectrum. Note that we define the absorption depth $d$ in terms of intensity attenuation $\exp(-d)$ through the sample. In Fig. \ref{Fig2}, we show absorption spectra taken with
the magnetic field parallel $\theta$=0$^{\circ}$ and at $\theta$=45$^{\circ}$ to the crystal
symmetry c-axis. In these measurements the light was polarized parallel to the c-axis for
maximum absorption \cite{Hastings2008}.

For the $\theta$=0$^{\circ}$ orientation, see Fig. \ref{Fig2}a, the b and c transitions are not resolved, but we can resolve two very weak a and d transitions at the expected positions. For this orientation we measured a branching ratio $R_{\|(exp)}\approx$0.05. Note that this value is very approximate because of the difficulty of measuring the strengths of the weak a and d transitions in Fig. \ref{Fig2}a (particularly considering the varying background). The theoretical model predicts a branching ratio $R_{\|(th)}$=0 for a magnetic field oriented along the crystal c-axis (see Sec. \ref{sec_theory}). The small contributions from the a and d transitions that are seen in Fig.\ref{Fig2}a could be due to a slight misalignment
of the magnetic field, or due to a small magnetic dipole character for the
$^{4}$I$_{9/2}($Z$_{1})\rightarrow$ $^{4}$F$_{3/2}($Y$_{1})$ transition which
does not vanish for this orientation. Indeed, the crystal field calculation
shows that some $J$-mixing occurs in the $^{4}$F$_{3/2}$ excited state with a
small contribution of the $^{4}$F$_{7/2}$ multiplet. The $^{4}$I$_{9/2}%
($Z$_{1})\rightarrow$ $^{4}$F$_{7/2}$ transition with $\Delta J=-1$ has a
magnetic dipole character.

For a magnetic field oriented at $\theta$=45$^{\circ}$ to the crystal c-axis we can clearly resolve all four transitions, see Fig. \ref{Fig2}b. Note that there is a slight intensity asymmetry in the a/d and b/c transitions, which is not expected. This is due to variations in the laser intensity during the frequency scan, although most of the variation was compensated for using the reference beam (see Sec. \ref{exp_descr}). From this spectrum we measured a branching ratio of $R_{\|(exp)}$=0.29$\pm$0.02 (see below on data evaluation), which is in close agreement with the theoretical value $R_{\|(th)}$=0.27 (see Sec. \ref{sec_theory}) for a magnetic field at $\theta$=45$^{\circ}$.

\begin{figure}[ptb]
\includegraphics[width=.50\textwidth]{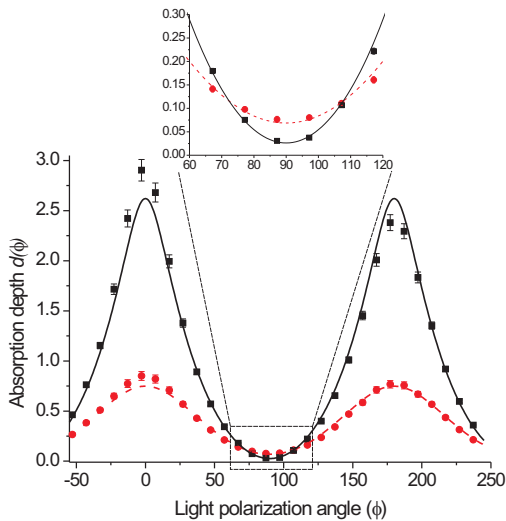}
\caption{Absorption depth $d(\phi)$ as a function of light polarization angle $\phi$ with respect to the crystal c-axis. Squares represent transitions b and c ($d_{bc}$), while circles represent transitions a and d ($d_{ad}$). We observe a clear inversion of the transition strength close to the $\phi$=90$^{\circ}$ orientation (see inset), as predicted by the theory (see text). The solid lines represent a fitted theoretical model explaining the variation in absorption as a function of angle. Note that the magnetic field orientation is $\theta$=45$^{\circ}$.}
\label{Fig3}
\end{figure}

The theoretical model also predicts that the branching ratio is inverted for a polarization orientated perpendicular to the crystal c-axis, i.e. $R_\parallel=1/R_\perp$. We investigated this by recording spectra while varying the light polarization angle. To extract the absorption coefficients we fit each spectrum to a simple model consisting of a sum of four Voigt functions, with (as free parameters) a common absorption depth for the b and c transitions $d_{bc}$, and for the a and d transitions $d_{ad}$. The absorption depths as a function of polarization angle are plotted in Fig. \ref{Fig3}.

The dependence of the absorption on the polarization angle can be understood by a simple model describing the anisotropic absorption. We write the incoming field vector (at $z$=0) as

\begin{eqnarray}
\mathbf{E}_0(\phi)&=&\mathbf{e}_\parallel E_\parallel(\phi) + \mathbf{e}_\perp E_\perp(\phi)\\ \nonumber &=&[ \mathbf{e}_\parallel \cos(\phi) + \mathbf{e}_\perp \sin(\phi)]E_0,
\label{polabseq1}%
\end{eqnarray}

\noindent where $\mathbf{e}_\parallel$ and $\mathbf{e}_\perp$ are unit vectors, $\phi$ is the polarization angle with respect to the c-axis, and $E_0$ is the field amplitude strength at $z$=0. The propagation of the field through the crystal is characterized by an anisotropic absorption and birefringence. It is important to note that the two orthogonal modes transform independently in the sample, due to the fact that the linear susceptibility is diagonal in the crystallographic axes for a crystal of tetragonal symmetry \cite{Boyd1992}. We can thus write the output field amplitude (at $z=L$) as

\begin{equation}
\mathbf{E}_L(\phi)=[\mathbf{e}_\parallel \cos(\phi) e^{-d_\parallel/2} + \mathbf{e}_\perp \sin(\phi) e^{-d_\perp/2}]E_0,
\label{polabseq2}%
\end{equation}

\noindent where we have neglected the relative phase factor induced by the birefringence. This we can do because we have no polarization dependent optics after the sample, where only the intensity is detected

\begin{eqnarray}
I_L(\phi) &=& [\cos^2(\phi) e^{-d_\parallel} + \sin^2(\phi) e^{-d_\perp}]I_0 \\ \nonumber &:=& e^{-d(\phi)}I_0.
\label{polabseq3}%
\end{eqnarray}

\noindent Here we have defined an effective polarization-dependent absorption depth $d(\phi)$. We fit this formula to the experimental data in Fig. \ref{Fig3} (shown as solid lines), with $d_\parallel$ and $d_\perp$ as free parameters. As seen the agreement is excellent for all polarization angles. Note that the weaker a and d transitions show a sinusoidal variation, while the stronger b and c transitions deviate from a sinus due to the exponential factors in the above formula. The fitted values are $d_{\parallel ad}$=0.75$\pm$0.02, $d_{\parallel bc}$=2.62$\pm$0.14, $d_{\perp ad}$=0.070$\pm$0.01 and $d_{\perp bc}$=0.025$\pm$0.01 (all errors are those from the fit shown in Fig. \ref{Fig3}). We can now calculate the branching ratios $R_\parallel$=$d_{\parallel ad}/d_{\parallel bc}$=0.29$\pm$0.02 and $R_\perp$=$d_{\perp ad}/d_{\perp bc}$=2.80$\pm$1.2. Our measurement confirms the theoretical prediction since $1/R_\perp$=0.36$\pm$0.15 which is equal to $R_\parallel$ within the experimental errors. It should be noted, however, that the error in $R_\perp $ is probably larger than above due to systematic errors when measuring absorptions at such a low level (some percent). Nevertheless, the expect inversion in the branching ratio is clearly observed.

To conclude this section, we have shown that it it possible to change the branching ratio by rotating the magnetic field with respect to the c-axis. While such studies have been carried out for non-Kramers ions \cite{louchet08}, we are not aware of similar studies for Kramers ions. Our experimentally measured branching ratios at angles $\theta$=0$^{\circ}$ and 45$^{\circ}$ agree well with the theoretical model presented in Sec. \ref{sec_theory}. A more detailed and quantitative comparison would, however, require more measurements as a function of angle. Nevertheless, our work show that a theoretical model is useful when choosing promising magnetic-field angles. Moreover, we observe experimentally the inversion of the branching ratio when changing the light polarization with respect to the c-axis.

\subsection{Optical pumping}

Here we will show that the $R$ value at the magnetic field orientation $\theta$=45$^{\circ}$ makes it possible to perform efficient optical pumping. In optical pumping experiments the branching ratio must have a non-zero value ($R > 0$) in order for the ions to have a certain probability of changing spin state while decaying from the excited state. How large it must be for the optical pumping to be efficient depends on the excited state lifetime and the ground state population lifetime \cite{lauritzen08}. We here show that the branching ratio at $\theta$=45$^{\circ}$ allows for much more efficient optical pumping than previous experiments at $\theta$=0$^{\circ}$ \cite{Hastings2008}.

The goal was to create a wide and deep transmission window
within the inhomogeneous absorption profile by optical pumping. The total pump
duration was 100 ms during which we made 1000 frequency sweeps over 20 MHz.
The resulting absorption profile was measured in the same way as before. The
probe pulses were much weaker in intensity in order not to modify the
absorption during the frequency scan. From the absorption profile the
percentage of ions left in the initial state could be estimated. The orientation of the polarization was $\phi$=20$^{\circ}$.

\begin{figure}[ptb]
\includegraphics[width=.50\textwidth]{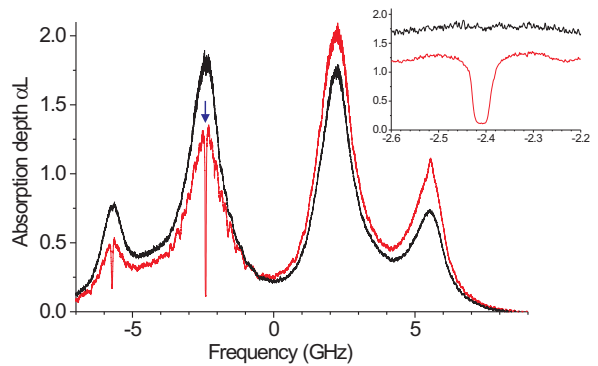}
\caption{Absorption spectra before (black line) and after (red line) optical
pumping. The optical pumping creates a 20 MHz wide transmission window in the
inhomogeneous absorption profile (indicated by the arrow). Note that the
pumping also creates another hole in the leftmost absorption peak, since this
transition starts from the same spin level. The pumped ions are moved into the
two right absorption peaks, where increased absorption is observed after
pumping. Inset: Zoom of the created transmission window.}
\label{Fig4}
\end{figure}

\begin{figure}[ptb]
\includegraphics[width=.50\textwidth]{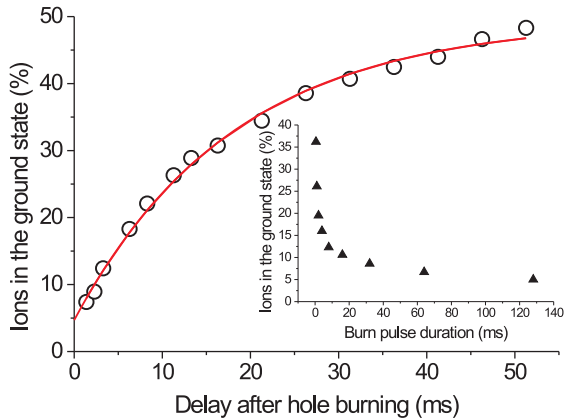}
\caption{Fraction of remaining ions in the pumped spin state as a function of
the delay time after the optical pumping. The solid line is a fit with a time constant of 18 ms. Extrapolating the fit to zero delay yields about 4 to 5 $\%$. Inset: Fraction of ions as a function of the pump pulse
duration. The highest pump efficiency is obtained for pulse durations longer
than 100 ms.}
\label{Fig5}
\end{figure}

In Fig. \ref{Fig4} we show the absorption profile measured 1.3 ms after the burning pulse. This delay allowed for all ions to decay back to the ground states (the excited state lifetime being $\sim$100$\mu$s). When optical pumping is applied (red line) a 20 MHz frequency window is created in the
absorption (no pumping is represented by the black line). In Fig. \ref{Fig5}
we show the percentage of ions left in the ground state, calculated from the
absorption spectra, as a function of the delay between the end of the pump
sequence and the time the scan reached the position of the bottom of the
created transmission window. When extrapolating the data to zero delay we find
that only $5\%$ of the ions are left in the initial state. In terms of the
total population, which is initially equally distributed over the two spin
states, only $2.5\%$ of the total population is left in the initial state.
This corresponds to a spin polarization of $97.5\%$. Note that these numbers
only apply to the ions that absorb within the spectral window created. We can
compare this optical pumping experiment to the one we made in a previous
work \cite{Hastings2008}, where we had the magnetic field oriented at
0$^{\circ}$. In that experiment as much as 15-20$\%$ of the ions were left in
the initial state (for a comparable magnetic field strength). We can thus
conclude that the optimization of the branching factor $R$ improved the
efficiency of the optical pumping by a factor 4-5.%

From the time-resolved absorption measurement, Fig. \ref{Fig5}, we found that
about $50\%$ of the initial absorption was recovered with a time constant of
18 ms. This decay time is given by spin population lifetime, since this decay
is much longer than the excited state lifetime of 100 $\mu$s. The remaining
$50\%$ of the initial absorption recovered with a significantly slower time
constant. Measurements at delay times of up to 800 ms gave a decay time of
about 320 ms and at these delays the absorption tends towards the original
absorption. It seems likely that this process is due to population trapping in
hyperfine states of the isotopes $^{143}$Nd and $^{145}$Nd (12$\%$ and 8$\%$
natural abundances, respectively). An isotopically pure crystal would,
however, be required in order to study this mechanism for population trapping.
Further improvement of the optical pumping process should also be possible by
higher cycling rate through stimulated emission on an auxiliary transition
\cite{lauritzen08}. Stimulated emission on for example the well-known 1064 nm
laser transition in Nd should decrease the excited state lifetime
considerably. In addition one could force spin-mixing of the excited state
Zeeman levels via radio-frequency excitation to artificially increase the
branching ratio \cite{lauritzen08}.

\section{Conclusions}

We have shown that Nd$^{3+}:$YVO$_{4}$ can provide an efficient
Zeeman-level $\Lambda$-system by tuning the magnetic field orientation. The branching ratio is extracted by measuring
the absorption in a moderate magnetic field where the inhomogeneously
broadened Zeeman transitions can be spectrally resolved. We found that the
branching ratio can be tuned by optimization of the angle between the external
magnetic field and the crystal symmetry c-axis. We find good agreement between a
theoretical model and our experimental data. \\ Finally we
used our findings about the branching ratio in order to improve spin preparation experiments via optical pumping in this material. We created a 20 MHz wide transmission window, where only 5 $\%$ of the ions were left in
the initial spin state. Further improvement of the optical pumping rate should
be possible with spin mixing and stimulated emission techniques.\newline In
addition to the findings presented here it is known that neodymium shows one
of the highest absorption of all rare-earth ions, a short excited state
lifetime of about 100 $\mu$s \cite{gui05} and good optical coherence properties
\cite{Sun2002,Hastings2008}. Furthermore stable diode lasers and efficient single
photon counters are available at the transition of interest in neodymium,
which makes this system interesting for various quantum information and
processing protocols.

\section{Acknowledgments}

This work was supported by the Swiss NCCR Quantum Photonics, the European Commission through the integrated project QAP, and the ERC Advanced Grant QORE.

\section{References}

\bibliographystyle{elsarticle-num}

\end{document}